\documentclass[12pt]{iopart}

\usepackage{amssymb,graphicx,epsfig}

\DeclareMathAccent{\bhat}{\mathalpha}{operators}{"5E}

\def\vk{\bi{k}}

\def\vE{\bi{E}}

\def\imag{{\rm i}}

\def\ket#1{|{#1}\rangle}
\def\bra#1{\langle{#1}|}

\def\threej#1#2#3#4#5#6{\left(\matrix{
				 #1&\;#2&\;#3 \cr
				 #4&\;#5&\;#6 }
				\right)}
\def\sixj#1#2#3#4#5#6{\left\{\matrix{
				 #1&\;#2&\;#3 \cr
				 #4&\;#5&\;#6 }
			    \right\}}

\begin{document}

%\title{The Hanle effect in hydrogen lines from unresolved,\break
%isotropic magnetic and electric fields}

\title[Hydrogen polarization in turbulent electric fields]%
{Scattering polarization of hydrogen lines
in the presence of turbulent electric fields} 

\author{R.~Casini and R.~Manso Sainz\footnote{Now at the 
Instituto de Astrof{\'\i}sica de Canarias, V{\'\i}a L\'actea s/n,
E-38200 La Laguna, Tenerife, Spain}}

\address{High Altitude Observatory,
National Center for Atmospheric Research\footnote{The National 
Center for Atmospheric Research is sponsored by the National 
Science Foundation}\break
P.O.~Box 3000, Boulder, CO 80307-3000, U.S.A.}

\ead{\mailto{casini@ucar.edu},\mailto{rsainz@iac.es}}

%\affiliation{P.O. Box 3000, Boulder, CO~80307-3000}

\begin{abstract}

We study the broadband polarization of hydrogen lines produced by 
scattering of radiation, in the presence of isotropic electric fields. 
In this paper, we focus on two distinct problems: a) the possibility 
of detecting the presence of turbulent electric fields by polarimetric 
methods, and b) the influence of such fields on the polarization due 
to a macroscopic, deterministic magnetic field. We found that isotropic 
electric fields decrease the degree of linear polarization in the
scattered radiation, with respect to the zero-field case. On the other 
hand, a distribution of isotropic electric fields superimposed onto 
a deterministic magnetic field can generate %determine 
a significant increase of 
the degree of magnetic-induced, net circular polarization. This 
phenomenon has important implications for the diagnostics of magnetic 
fields in plasmas using hydrogen lines, because of the 
ubiquitous presence of the Holtsmark, microscopic electric field
from neighbouring ions. In particular, previous 
solar magnetographic studies of the Balmer lines of hydrogen may need 
to be revised because they neglected the effect of turbulent electric 
fields on the polarization signals. In this work, we %only present
give explicit 
results for the Lyman~$\alpha$ and Balmer~$\alpha$ lines.
%of these lines. 
%We also discuss the interest of these results 
%for the diagnostics of electric microfields (and hence, of ionic 
%densities) in a plasma, as well as of macroscopic, unresolved electric 
%fields. 
\end{abstract}

\pacs{32.60.+i, 32.80.Bx, 52.25.Os}

\maketitle

\begin{table}
\caption{\label{tab:table}
Holtsmark normal field strengths of microturbulent electric fields in
hydrogen plasmas of laboratory and astrophysical interest.}
\begin{indented}
%\item[]\begin{tabular}{@{}1111}
\centering
\vspace{5pt}
\item[]
\begin{tabular}{cccc}
\br
%\hline\hline
Object & $N\,\rm(cm^{-3})$ & $E\,\rm (V\,cm^{-1})$ & $E\,\rm (e.s.u.)$ \\
\mr
%\hline
Interstellar gas & 1 & 3.75 10$^{-7}$ & 1.25 10$^{-9}$ \\
% & 10$^1$ & 1.74 10$^{-6}$ & 5.80 10$^{-9}$ \\
% & 10$^2$ & 8.08 10$^{-6}$ & 2.69 10$^{-8}$ \\
Gaseous nebula & 10$^3$ & 3.75 10$^{-5}$ & 1.25 10$^{-7}$ \\
% & 10$^4$ & 1.74 10$^{-4}$ & 5.80 10$^{-7}$ \\
% & 10$^6$ & 3.75 10$^{-3}$ & 1.25 10$^{-5}$ \\
% & 10$^8$ & 8.08 10$^{-2}$ & 2.69 10$^{-4}$ \\
Solar Corona & 10$^9$ & 3.75 10$^{-1}$ & 1.25 10$^{-3}$ \\
Solar Corona, solar transition region & 10$^{10}$ & 1.74 & 5.80 10$^{-3}$ \\
Solar Chromosphere, solar prominences & 10$^{11}$ & 8.08 & 2.69 10$^{-2}$ \\
Solar $T$ minimum, diffuse hot plasma & 10$^{12}$ & 3.75 10$^{1}$ & 1.25 10$^{-1}$ \\
Solar Photosphere, gas discharge & 10$^{14}$ & 8.08 10$^{2}$ & 2.69 \\
Thermonuclear plasma & 10$^{15}$ & 3.75 10$^{3}$ & 1.25 10$^{1}$ \\
Dense hot plasma & 10$^{18}$ & 3.75 10$^{5}$ & 1.25 10$^{3}$ \\
Laser plasma & 10$^{20}$ & 8.08 10$^{6}$ & 2.69 10$^{4}$ \\
% & 10$^{25}$ & 1.74 10$^{10}$ & 5.80 10$^{7}$ \\
\br
%\hline \hline
\end{tabular} 
\end{indented} 
\end{table}

\section{Introduction} \label{sec:intro}

Radiation scattering by a plasma is due to the processes of photon 
absorption and re-emission by the atoms in the plasma. During 
absorption, atomic polarization (i.e., population imbalances and 
quantum coherences between atomic sublevels) can be created, depending 
on the angular distribution and polarization of the incident radiation 
(\textit{optical pumping\/}).

In the presence of external fields, the natural degeneracy of the 
angular-momentum sublevels is removed, so these sublevels evolve at 
slightly different frequencies. As a consequence, the atomic 
polarization---and therefore the polarization of the re-emitted
radiation---is modified in a way that depends on the strength and 
orientation of the external field (e.g., by the Hanle effect). It is 
noteworthy 
that the scattering polarization does not necessarily cancel out even if 
the external fields are isotropically distributed, unless the atomic 
excitation processes (either radiative or collisional) are also isotropic. 
The phenomenon of (partial) depolarization of the scattered radiation 
due to isotropically distributed magnetic fields has been extensively 
investigated \cite{ST82,LL88,ST94,LL04}. In the case of hydrogen, 
because of its sensitivity to electric fields to first order of 
perturbation (\textit{linear Stark effect\/}), we must expect that 
the line polarization due to radiation scattering be significantly 
affected by both magnetic and electric fields. 

This suggests two interesting problems, which are related. The first 
problem concerns the possibility of detecting turbulent electrostatic 
fields by polarimetric methods. The fields can be either microscopic 
(due to neighboring ions) or macroscopic (due to plasma turbulence)
\cite{GR74}. The second problem concerns the effect of turbulent 
electric fields on the polarization signals generated in the presence of
a deterministic magnetic field, and the consequences for the diagnostics 
of macroscopic magnetic fields using hydrogen lines.
%%%%%
%We shall show that ``standard'' diagnostics and magnetograms in the
%Ha and Lya ignoring these effecst may be seriously flawed.

Table~\ref{tab:table} lists approximate values for the strength, $E$, of a 
microscopic electric field acting on an atom as a function of the 
perturbing ion density, $N$, in the plasma. These values are calculated 
as the Holtsmark normal field strength characteristic of a random 
distribution of neighboring ions \cite{GR74}. They illustrate the limit 
of the electric contribution to scattering polarization that must be 
expected under different plasma conditions.

%An estimation of the field may be obtained assuming that random 
%fluctuations of density yield an imperfect charge cancellation 
%equivalent to the field produced by an elementary charge $e$ at a mean 
%distance $r=N^{-1/3}$.
%Therefore, $E=e/(4\pi\varepsilon_0 r)$ ($\varepsilon_0$ being the dielectric
%constant of vacuum). A more precise approximation is 2.66 times this
%value (REF), is given in columns 3 and 4 of Table 1.  

%\begin{tab:table}
%\begin{tabular}{ccccc}
%\hline\hline
%Object & N (cm$^{-3}$) & E (V m$^{-1}$) & E (V\,cm$^{-1}$) & E (e.s.u.) \\
%\hline
% & 10 & 6.68 10$^{-5}$ & 6.68 10$^{-7}$ & 2.23 10$^{-9}$ \\
% & 10$^2$ & 3.10 10$^{-4}$ & 3.10 10$^{-6}$ & 1.03 10$^{-8}$ \\
% & 10$^4$ & 6.68 10$^{-3}$ & 6.68 10$^{-5}$ & 2.23 10$^{-7}$ \\
% & 10$^6$ & 1.44 10$^{-1}$ & 1.44 10$^{-3}$ & 4.80 10$^{-6}$ \\
% & 10$^8$ & 3.10 10$^{0}$ & 3.10 10$^{-2}$ & 1.03 10$^{-4}$ \\
% & 10$^{10}$ & 6.68 10$^{0}$ & 6.68 10$^{-1}$ & 2.23 10$^{-3}$ \\
%Solar Chromosphere/ Prominence & 10$^{11}$ & 3.10 10$^{2}$ & 3.10 10$^{0}$ & 1.03 10$^{-2}$ \\
%Solar T minimum & 10$^{12}$ & 1.44 10$^{3}$ & 1.44 10$^{1}$ & 4.80 10$^{-2}$ \\
%Solar Photosphere & 10$^{15}$ & 1.44 10$^{5}$ & 1.44 10$^{3}$ & 4.80 10$^{0}$ \\
% & 10$^{20}$ & 3.10 10$^{8}$ & 3.10 10$^{6}$ & 1.03 10$^{4}$ \\
% & 10$^{25}$ & 6.68 10$^{11}$ & 6.68 10$^{9}$ & 2.23 10$^{7}$ \\
%\hline \hline
%\end{tabular}
%\end{tab:table}

In Section~\ref{sec:turbulent}, we will study the scattering polarization 
of hydrogen lines in the presence of isotropically distributed magnetic or 
electric fields. In Section~\ref{sec:Eturbulent}, we investigate instead 
the case in which the scattering polarization due to a deterministic 
magnetic field of various orientations is modified by the additional 
presence of an isotropic distribution of electric fields. In this work
we only present results for the Lyman~$\alpha$ (Ly$\alpha$) 
and Balmer~$\alpha$ (H$\alpha$) lines, although we verified
the generality of our conclusions for all hydrogen transitions up to 
level $n=4$.

\begin{figure}[t]
\flushright \leavevmode
\includegraphics[scale=.75]{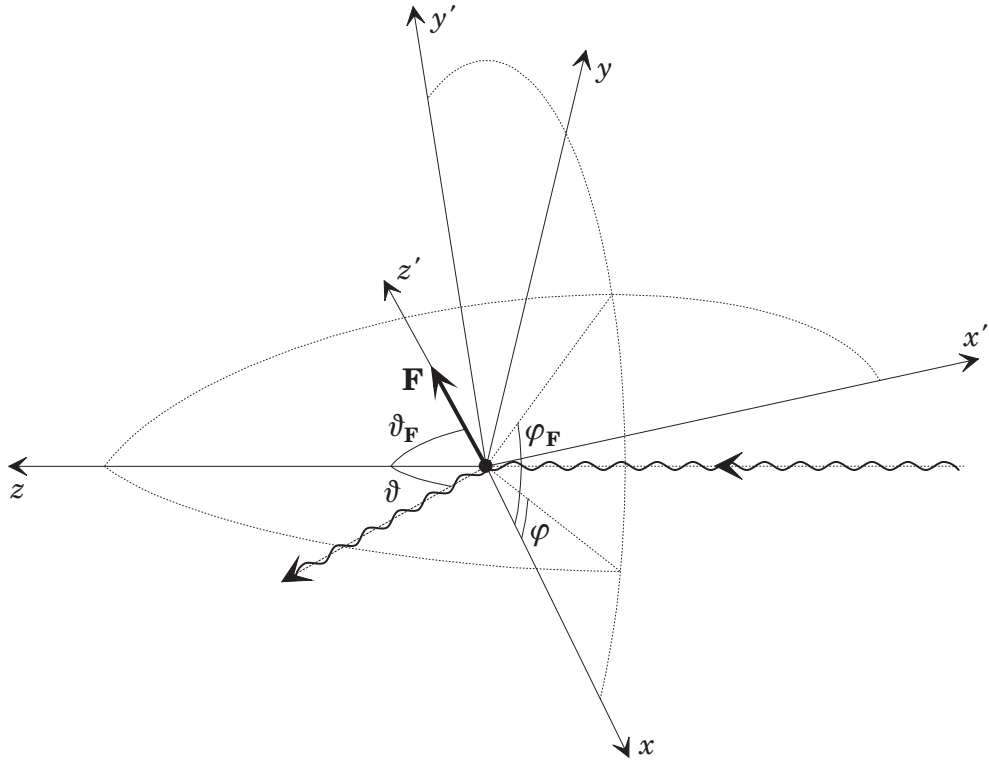}
\caption{\label{fig:geometry}
General geometry for the description of the scattering of radiation 
in the presence of an external field. The pair $(\vartheta_\bi{F},\varphi_\bi{F})$ 
specifies the direction of the external field, $\bi{F}$, in the
reference frame $S\equiv xyz$, whereas the pair $(\vartheta,\varphi)$ 
specifies the direction of the scattered radiation.
The reference frame $S'\equiv x'y'z'$ is the one adopted for the solution 
of the statistical equilibrium of the atomic system subject to the external 
field and to the incident radiation along $z$, and it is obtained from
the original frame $S$ by means of a rotation of $\varphi_\bi{F}$ around $z$
plus a rotation of $\vartheta_\bi{F}$ around $y'$.}
\end{figure}

\section{Scattering polarization in the presence of randomly oriented fields}
\label{sec:turbulent}

%The most general description of the excitation state of hydrogen atoms
%in the presence of electric and magnetic fields is provided by
%the density matrix components $\langle nS L J M_J|\boldsymbol{\rho}| nS L' J' M_{J'}\rangle$.
%Equivalent and more convenient for our problem, because of their 
%natural spatial rotation properties, is a description in terms 
%of the irreducible spherical components ${}^{nS}\rho^K_Q(LJ,L'J')$ (REF).

The general expression for the emissivity in the four Stokes parameters
\cite{BW83,LL04}, %for their definition),
$I$, $Q$, $U$, and $V$, %(see 
for a gas of hydrogen atoms subject to electric and magnetic fields, was 
given in \cite{CA05}, assuming $L$-$S$ coupling. Here we %only 
consider broadband 
(i.e., frequency-integrated) polarization signals, in which case the line 
emissivity for a transition from an upper level of principal quantum
number $n_u$ to a lower level $n_\ell$, in a gas of atomic density $N$, 
is \cite{CA05}
\begin{eqnarray} \label{eq:emissivity}
\fl
\varepsilon_i(\vk)
&=&\frac{e_0^2 a_0^2}{\sqrt{3}\,\pi c^3}\,N
	\omega_{n_un_l}^4 \sum_{L_uL_u'L_\ell}\sum_{J_uJ_u'}\sum_{KQ}
	(-1)^{L_u+L_u'+L_\ell+S+J_u'+K+1} \nonumber \\
\fl
&&\times 
	\Lambda(n_u L_u, n_\ell L_\ell) \Lambda(n_uL'_u, n_\ell L_\ell)
	\sqrt{(2 J_u+1)(2 J'_u+1)} \nonumber \\
\fl
&&\times
	\sixj{L_u}{L'_u}{K}{1}{1}{L_\ell}
	\sixj{L_u}{L'_u}{K}{J'_u}{J_u}{S}
	T^K_Q(i, \vk)\,{}^{n_uS}\rho^K_Q(L_u J_u,L_u'J_u')\;,
\end{eqnarray}
where $i=0,1,2,3$ enumerates the four Stokes parameters. In 
eq.~(\ref{eq:emissivity}), $\omega_{n_u n_\ell}$ is the Bohr
frequency of the transition, and
\begin{displaymath}
\Lambda(nL,n'L')=\sqrt{(2L+1)(2L'+1)}\,\bra{nL}r\ket{n'L'}
\threej{L}{L'}{1}{0}{0}{0}\;,
\end{displaymath}
where the dipole-matrix element is calculated using Gordon's formula 
\cite{BS57}.
The angular distribution of the emitted radiation is determined by 
the irreducible spherical tensors $T^K_Q(i,\vk)$ ($K=0,1,2$, and
$Q=-K,\dots,K$), where $\vk$ is the propagation direction. The explicit 
expressions of these tensors are tabulated, for example, by 
\cite{BO97,LL04,CA05}.
Figure~\ref{fig:geometry} illustrates the relevant geometric quantities
for the description of radiation scattering in the presence of an 
external field.

For a given illumination, the irreducible spherical components of the 
density matrix, ${}^{nS}\rho^K_Q(L_u J_u,L_u'J_u')$, are calculated 
in the limit of statistical equilibrium, taking into account the relaxation
and transfer of atomic polarization between levels, because of absorption and 
emission processes, as well as the level mixing induced by the external 
fields \cite{CA05}. 
%\textbf{
In particular, the self-consistent treatment
of level mixing within our formalism allows us to consider all possible
regimes of magnetic and electric fields in a unified way, including the
magnetic and electric Hanle effects, the linear Zeeman and Stark effects
in both the strong- and weak-field limit (i.e., both ``normal'' and
``anomalous'' effects), as well as the incomplete Paschen-Back effect 
for both magnetic and electric fields (which characterizes the actual 
regimes for level crossing).%}
It is important to notice that magnetic fields only 
mix atomic sublevels with the same $L$, whereas electric fields only mix 
sublevels with $\Delta L=\pm 1$. For arbitrary field geometries, both fields 
mix sublevels with different $J$'s and $M$'s. Thus, in the 
presence of both electric and magnetic fields, only the principal quantum 
number, $n$, remains a good quantum number.\footnote{We limit our 
investigation to field strengths for which configuration mixing between 
different $n$ levels can be neglected.} We will consider
an atomic model of hydrogen including the Bohr levels $n=1,2,3,4$.
Therefore, the excitation state of the atomic system is described by
$4+64+324+1024=1416$ density matrix elements 
${}^{nS}\rho^K_Q(L_u J_u,L_u'J_u')$.
% elements 
%are required to describe The statistical equilibrium equations for the
%first four $n$ levels
These form the solution of a 1416$\times$1416 linear system, which must
be solved numerically. 
%\textbf{
In this solution, all allowed radiative
transitions within the atomic model are considered simultaneously and
consistently. So the contributions of both elastic (Rayleigh) and inelastic 
(Raman) scattering to the broadband polarization of the scattered
radiation are properly taken into account.%}
%This system is made non singular 
%by substituting one redundant equation (that for the population of the
%ground level), by that for the conservation of particles. 

To investigate the effects of isotropically distributed fields, we must
calculate the average $\langle \varepsilon_i(\vk)
\rangle_{(\vartheta_\bi{F},\varphi_\bi{F})}$ over all possible field 
orientations $(\vartheta_\bi{F},\varphi_\bi{F})$ within the sphere. 
%Since $\varepsilon_i(\vk)$ 
%is linear in $\rho^K_Q(JL,J'L')_{(\vartheta_\bi{F},\varphi_\bi{F})}$, this is 
%equivalent to calculating $\varepsilon_i(\vk)$ for the averaged
%$\langle \rho^K_Q(JL,J'L') \rangle_{(\vartheta_\bi{F},\varphi_\bi{F})}$. 
%Because of the relatively large algebraic system involved, this can 
%only be done numerically. 
In practice, we compute $\varepsilon_i(\vk)$ 
for a finite number of directions, ($\vartheta_\bi{F}^i, \varphi_\bi{F}^j$),
determined as the nodes of some angular quadrature over the sphere, 
and then we sum the results with the appropriate weights $w_{i,j}$ 
given by the quadrature formula.

\begin{figure}[t]
\flushright \leavevmode
\includegraphics[scale=.4]{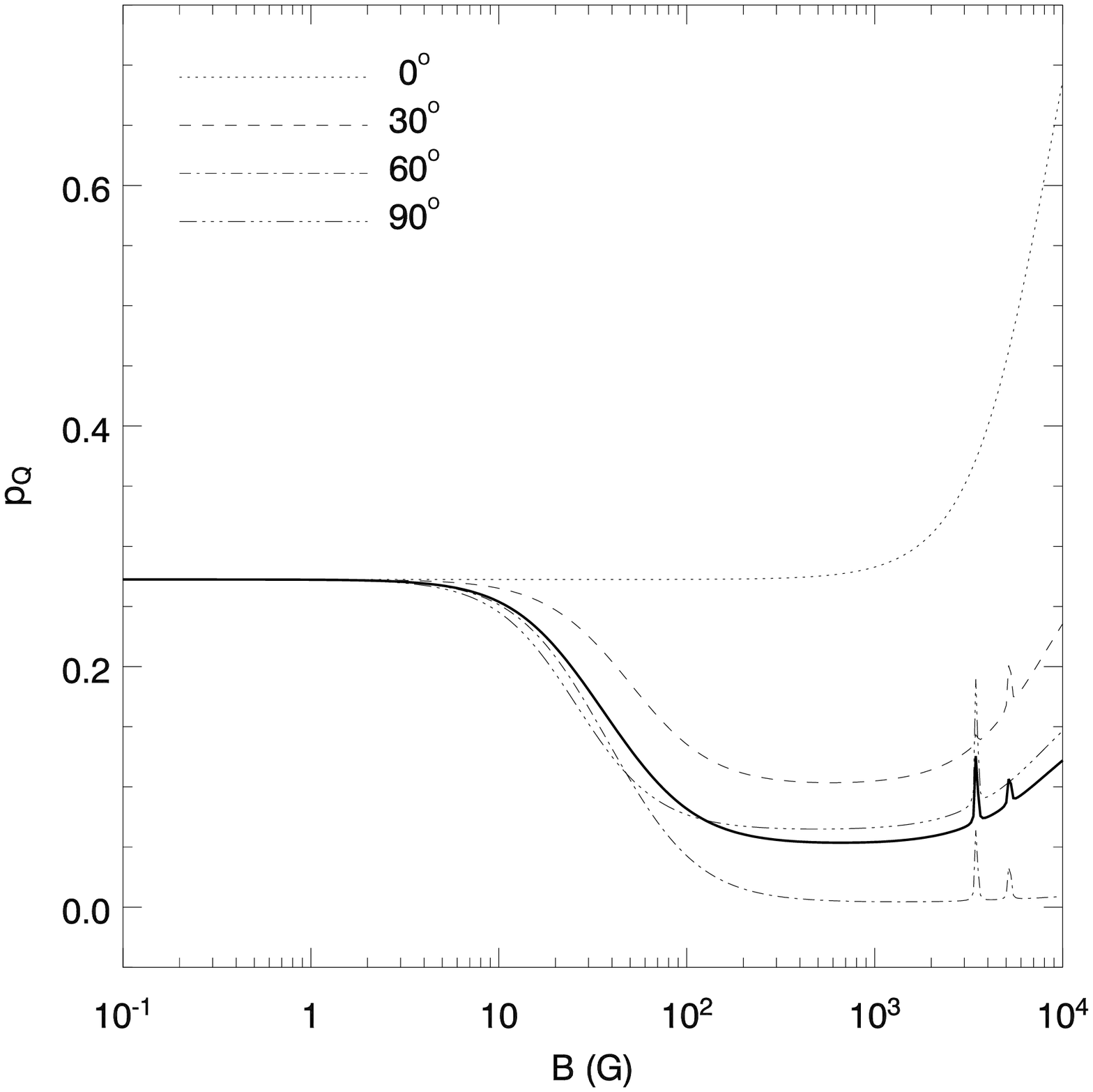}\hspace{5pt}
\includegraphics[scale=.4]{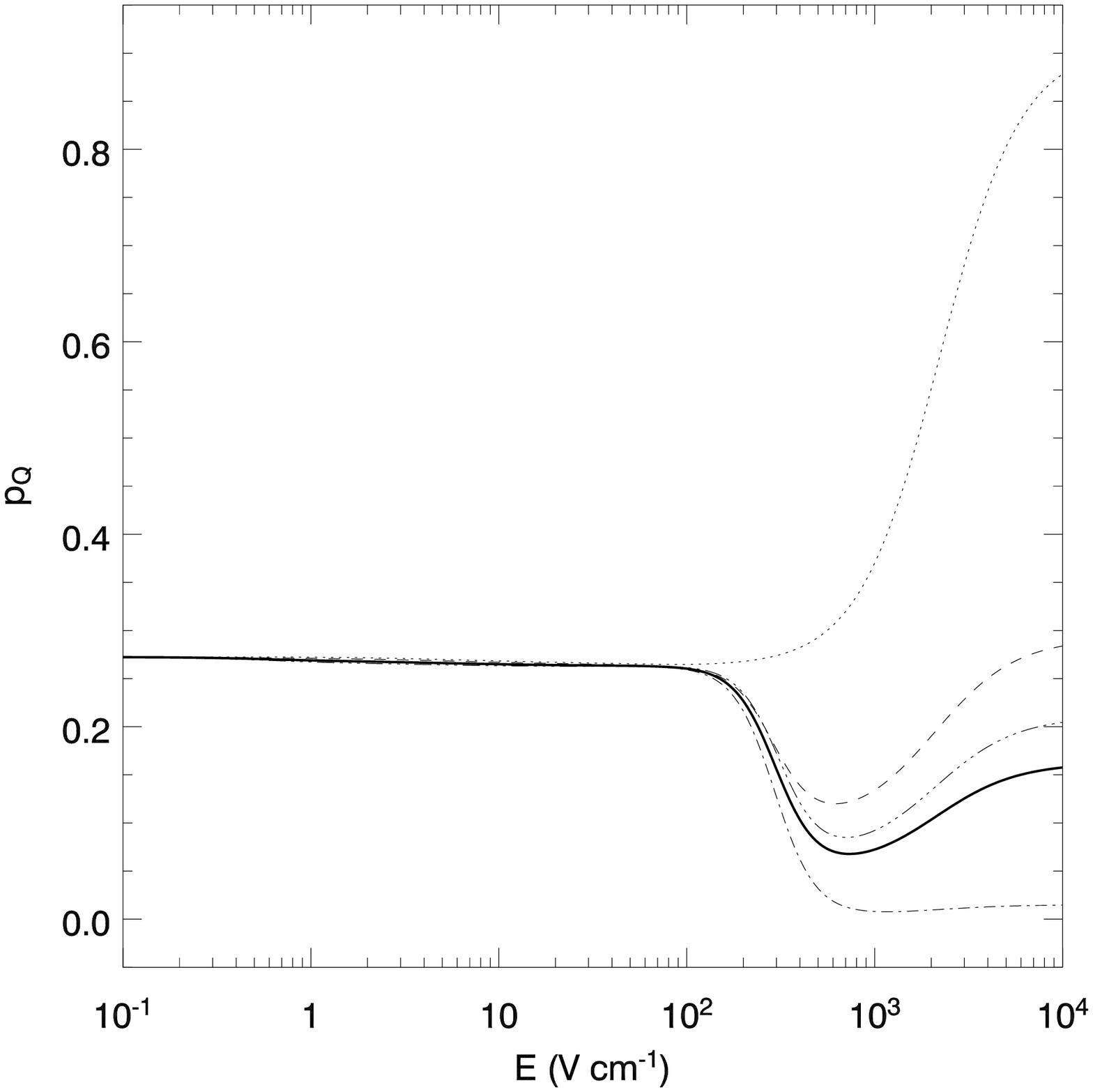}
\caption{\label{fig:Lyman} %\small
Broadband, fractional linear polarization of the 
Ly$\alpha$ line of hydrogen in a $90^\circ$-scattering event, as a 
function of the strength of a randomly oriented magnetic field 
(left panel) and electric field (right panel). The thick solid line 
corresponds to the case of isotropic fields, while the discontinuous 
lines show the contributions of azimuth-averaged fields of various 
inclinations $\vartheta_\bi{F}$. Because of the symmetry properties of 
the scattering process, the curves for $\vartheta_\bi{F}$ and 
$180^\circ-\vartheta_\bi{F}$ are identical.}
\end{figure}

For the case considered in this section, it is possible to perform
the analytic integration over the field azimuth, $\varphi_\bi{F}$, for any
given value of the field inclination, $\vartheta_\bi{F}$, because 
of the axial symmetry of the problem around the propagation direction 
of the incident radiation. In order to see this, we consider in detail the
form of eq.~(\ref{eq:emissivity}). The geometric tensors, $T^K_Q(i,\vk)$, 
are expressed in the reference frame that has the field direction as its 
polar axis (which we indicate with $z'$; see Fig.~\ref{fig:geometry}).
They are obtained from the same tensors in the reference frame specified 
by the incident radiation via the transformation \cite{BS93}
\begin{equation}
\label{eq:trans}
T^K_Q(i,\vk)=\sum_{Q'} D^K_{Q'Q}(R_{zz'})\,T^K_{Q'}(i,\vk)_z\;,
\end{equation}
where $R_{zz'}$ is the rotation operator that carries the $z$-frame into
the $z'$-frame. The rotation matrix in eq.~(\ref{eq:trans}) is 
\begin{equation}
\label{eq:rotmat}
D^K_{Q'Q}(R_{zz'})=\exp(-\imag Q'\varphi_\bi{F})\,d^K_{Q'Q}(\vartheta_\bi{F})\;,
\end{equation}
since the third Euler angle of the transformation can arbitrarily be set
equal to zero in this case. The azimuth of the scattering direction in 
the $z$-frame, $\varphi$, enters the expression of the geometric tensors 
$T^K_Q(i,\vk)_z$ in such a way that it is possible to factor it out, 
since (see Table~I in \cite{CA05})
\begin{equation}
\label{eq:factor}
T^K_Q(i,\vk)_z=t^K_Q(i,\vk)\,\exp(\imag Q\varphi)\;,
\end{equation}
where the tensors $t^K_Q(i,\vk)$ no longer depend on $\varphi$.
The azimuthal average of $T^K_Q(i,\vk)$ is then
(cfr.~eq.~[\ref{eq:trans}])
\begin{eqnarray} \label{eq:random_average}
\fl
\Bigl\langle T^K_Q(i,\vk)\Bigr\rangle_\varphi
&=&\sum_{Q'} d^K_{Q'Q}(\vartheta_\bi{F})\,t^K_{Q'}(i,\vk)\,
   \Bigl\langle \exp\bigl[\imag Q'(\varphi-\varphi_\bi{F})\bigr]
	\Bigr\rangle_\varphi
	=d^K_{0Q}(\vartheta_\bi{F})\,t^K_0(i,\vk) 
	\nonumber \\
\fl
&\equiv& d^K_{0Q}(\vartheta_\bi{F})\,T^K_0(i,\vk)_z\;.
\end{eqnarray}
From a physical point of view, averaging eq.~(\ref{eq:emissivity}) 
over the azimuth of the scattering direction for a given magnetic field 
must be equivalent to averaging instead over the azimuth of the magnetic 
field for a fixed scattering direction. Therefore, using the averages 
(\ref{eq:random_average}) in eq.~(\ref{eq:emissivity}), in place of the 
original tensors $T^K_Q(i,\vk)$, is equivalent to computing 
the scattered radiation from an elemental volume of plasma embedded in 
a field of random azimuth $\varphi_\bi{F}$.

\begin{figure}[t]
\flushright \leavevmode
\includegraphics[scale=.4]{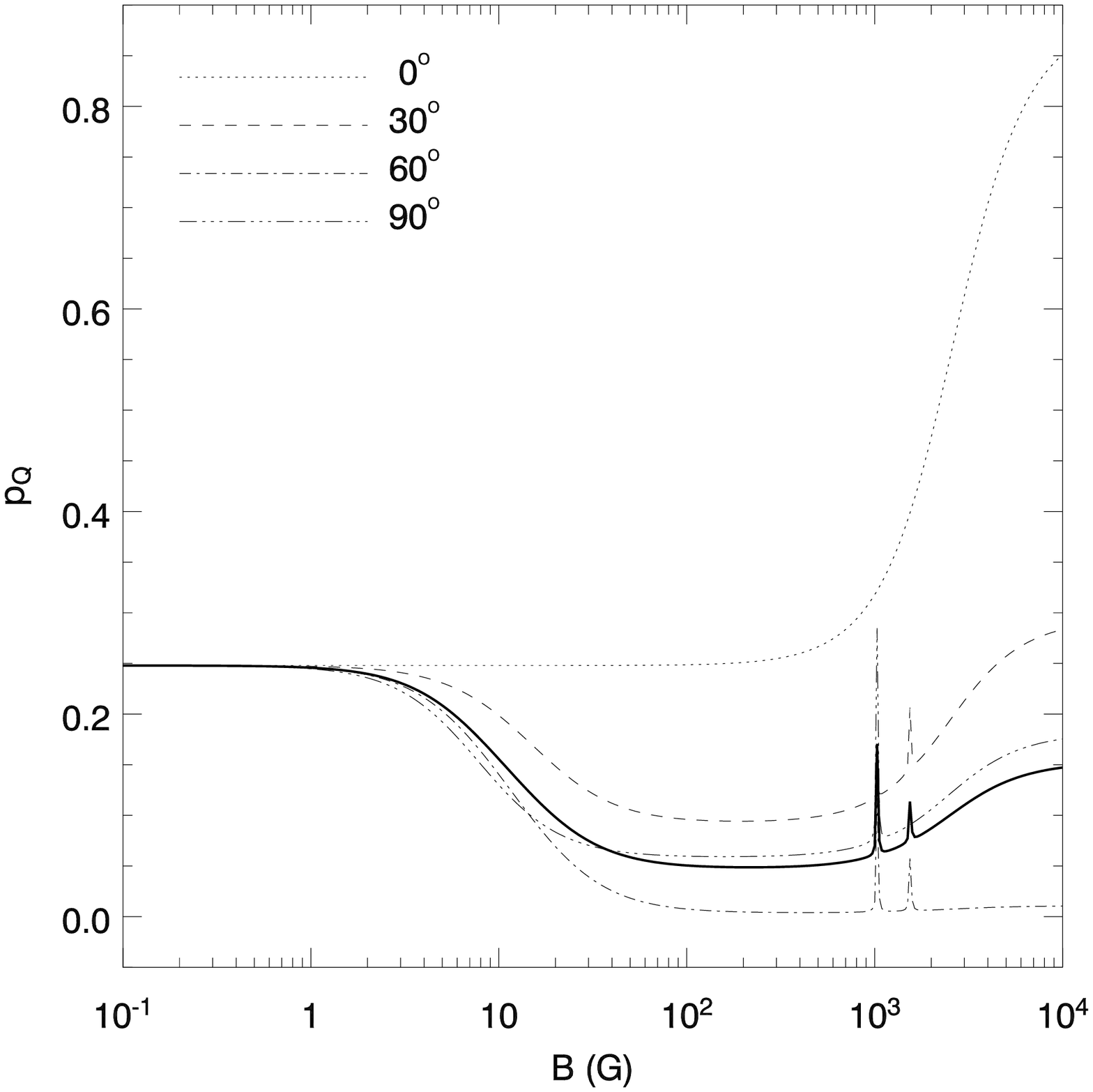}\hspace{5pt}
\includegraphics[scale=.4]{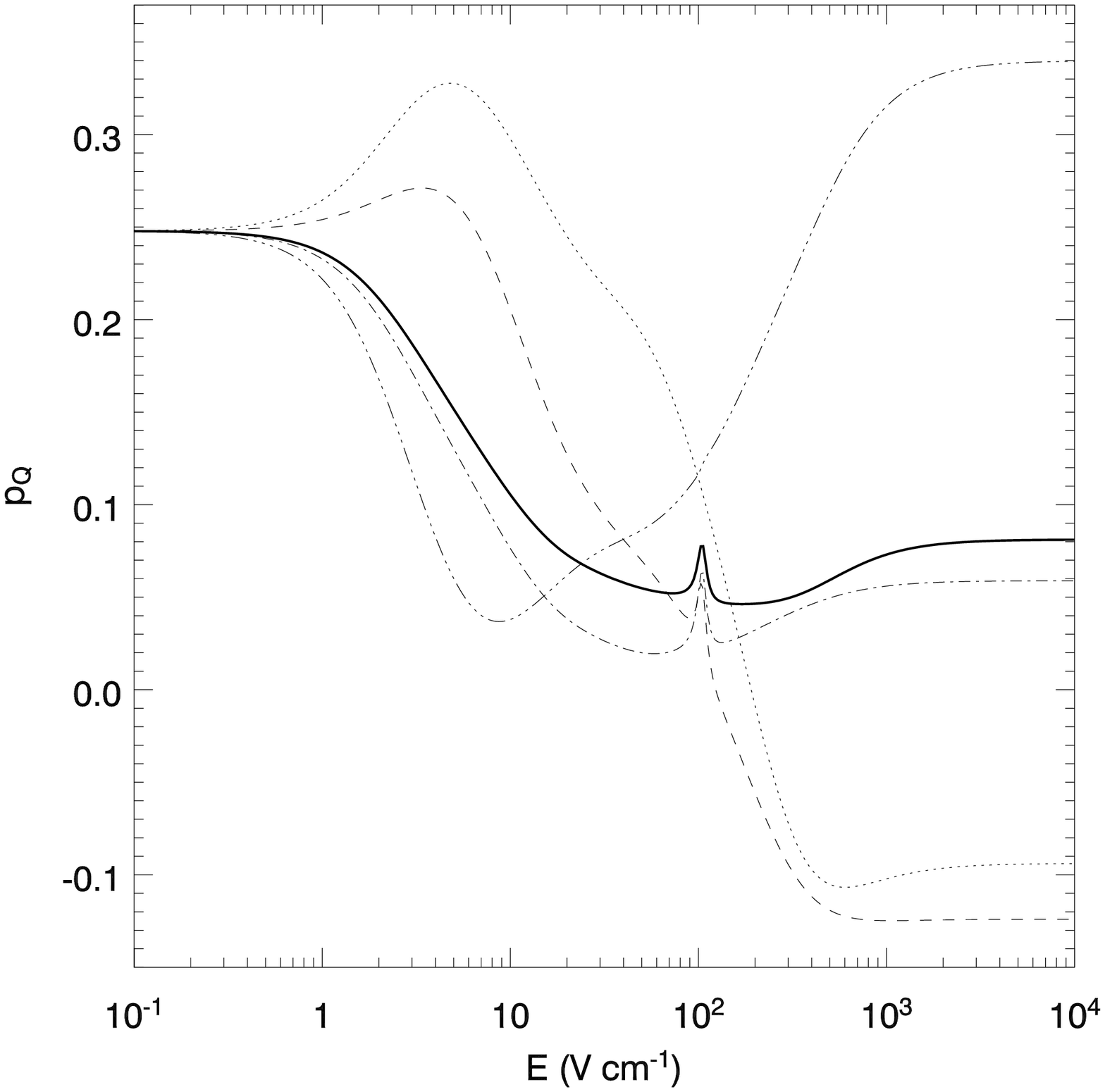}
\caption{\label{fig:Balmer} %\small
Same as Fig.~\ref{fig:Lyman}, but for the H$\alpha$ line
of hydrogen.}
\end{figure}

Figures~\ref{fig:Lyman} and \ref{fig:Balmer} show the broadband,
fractional linear polarization, $p_Q$, of the Ly$\alpha$ and 
H$\alpha$ lines of hydrogen (at $121.5\,\rm nm$ and $656.3\,\rm nm$, 
respectively) in a $90^\circ$-scattering event, as a function of 
the field strength of randomly oriented magnetic and electric 
fields. (Because of the symmetry of the problem, $p_U\equiv 0$, 
and therefore $p_Q$ contains the entire information of linear
polarization.) In those figures, we show the case of fields that 
are isotropically 
distributed in azimuth for various inclinations $\vartheta_\bi{F}$ 
from the $z$-axis (discontinuous lines), and the case of fields that 
are instead completely 
isotropic within the unit sphere (continuous thick line). The case of 
completely isotropic fields is calculated using an 8-point Gaussian 
quadrature of the solutions for random-azimuth fields at appropriate 
inclinations \cite{AS64}. Because of the symmetry 
properties of the Hanle effect in $90^\circ$ scattering, it is possible 
to consider only the hemisphere with $\vartheta_\bi{F}\le 90^\circ$, since 
the polarization curves for $\vartheta_\bi{F}$ and for 
$180^\circ-\vartheta_\bi{F}$ are identical. For the calculations of these 
plots, we assume that the atom is illuminated by a collimated beam of 
radiation corresponding to a Planckian intensity at $20\,000\,\rm K$. 
%and the atomic model of hydrogen we adopt includes all Bohr levels up 
%to $n=4$ \cite{CA05}.

We see that random-azimuth fields can produce either an increase 
or a decrease of the broadband linear polarization with respect 
to the field-free scattering polarization, depending on the field
strength, and also on the field inclination with respect to the direction 
of incident radiation. However, the overall trend for isotropically 
distributed fields within the sphere is always one of depolarization 
of the scattered radiation.

The strong resonances shown in the left panels of Figs.~\ref{fig:Lyman} 
and \ref{fig:Balmer} correspond to different crossings between magnetic 
sublevels within the upper levels of the transitions. In a $\rm {}^2P$ 
term, the two sublevels $M=+1/2$ and $M=-1/2$ of the $\rm {}^2P_{1/2}$ 
level cross the sublevel $M=-3/2$ of the $\rm {}^2P_{3/2}$ levels
respectively at
$\omega_B=(4/9)\,\omega_{\rm FS}$ and 
$\omega_B=(2/3)\,\omega_{\rm FS}$, where $\omega_B$ is the Larmor
frequency for the applied field, and $\omega_{\rm FS}$ is the
fine-structure separation between the two $J$ levels of the term
\cite{LL04,BS57}. In the particular case of hydrogen, these
crossings occur at $B=3483\,\rm G$ and $B=5225\,\rm G$ in the 
$\rm 2\,{}^2P$ term, and at $B=1032\,\rm G$ and $B=1548\,\rm G$ in
$\rm 3\,{}^2P$ term. Because at level crossing the atomic system 
tends to recover the polarization state that characterizes the
absence of Hanle depolarization (corresponding to the curves 
for vertical magnetic fields in Figs.~\ref{fig:Lyman} and 
\ref{fig:Balmer}), these level-crossing resonances produce important 
peaks in the scattering polarization signal 
\cite{BO80,MS90}. Because in a $90^\circ$-scattering event only 
the $\Delta M=\pm2$ crossings contribute to scattering 
polarization when $\vartheta_\bi{B}=90^\circ$, the resonance for 
$\omega_B=(2/3)\,\omega_{\rm FS}$ disappears from our plots for 
such inclination of the field \cite{BO80}. 

The resonance of the H$\alpha$ polarization for $E=104.5\,\rm V cm^{-1}$,
visible in the right panel of Fig.~\ref{fig:Balmer}, has an analogous 
physical explanation. It corresponds to the crossing between pairs of 
sublevels with $|M|=1/2$ and $|M|=3/2$ that are attributable to the 
$\rm 3\,{}^2P_{3/2}$ term in the limit of vanishing fields.

It must be noted that the effects of 
level crossing take place in an interval of Larmor frequencies
(magnetic or electric) comparable in width to the inverse of the 
lifetime of the levels 
involved. Thus the width of the associated resonances must also be
comparable to that of the interval within which the Hanle-effect 
depolarization (seen as a gentle slope immediately following the
initial plateau in Figs.~\ref{fig:Lyman} and \ref{fig:Balmer}) 
takes place. Since the level-crossing resonances occur at much 
larger Larmor frequencies (comparable to the fine-structure 
separation) than the Hanle depolarization, they appear as very 
sharp peaks in our figures, because of the logarithmic 
scale adopted for the field strength.

The depolarization of hydrogen line radiation in the presence of isotropic 
electric fields has important consequences for the diagnostics of
magnetic fields in laboratory and astrophysical plasmas.
For example, it is customary in solar plasma diagnostics to assume
that depolarization of the scattered radiation is indicative of the
magnetic Hanle effect (and possibly, of collisional depolarization), 
and therefore it is used to infer the magnetic field in the plasma. 
The fact that an isotropic distribution of electric fields as low as 
$10\,\rm V\,cm^{-1}$ (i.e., a Holtsmark field typical of the plasma 
density of the solar atmosphere; see Table~\ref{tab:table}) can be 
largely responsible for the depolarization of the scattered radiation 
of the H$\alpha$ line of hydrogen is a result that must be taken into 
consideration for any modeling of the magnetic Hanle effect of this 
commonly observed line. 

\begin{figure}[t]
\flushright \leavevmode
\includegraphics[height=3in]{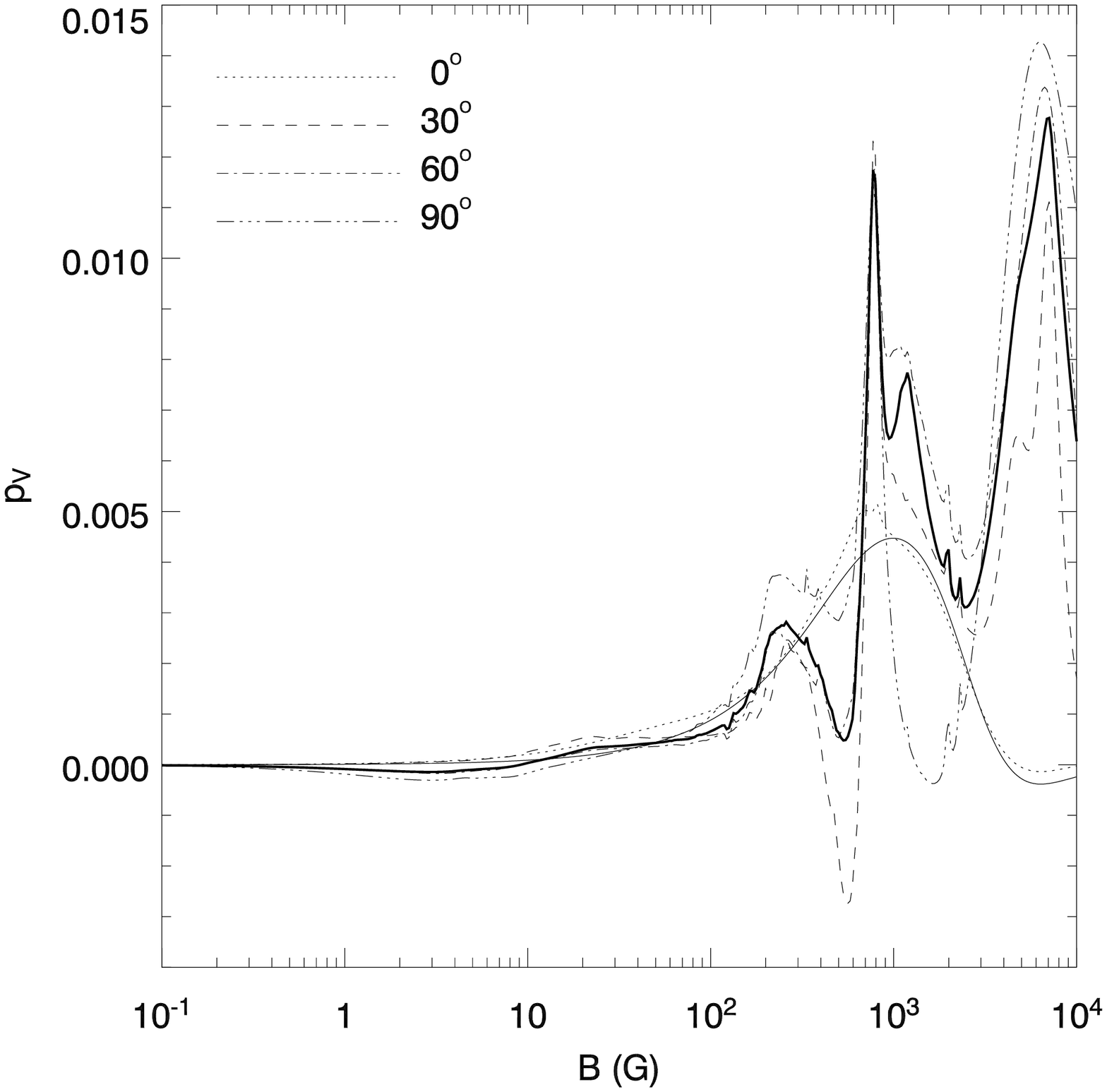}\hspace{5pt}
\includegraphics[height=3in]{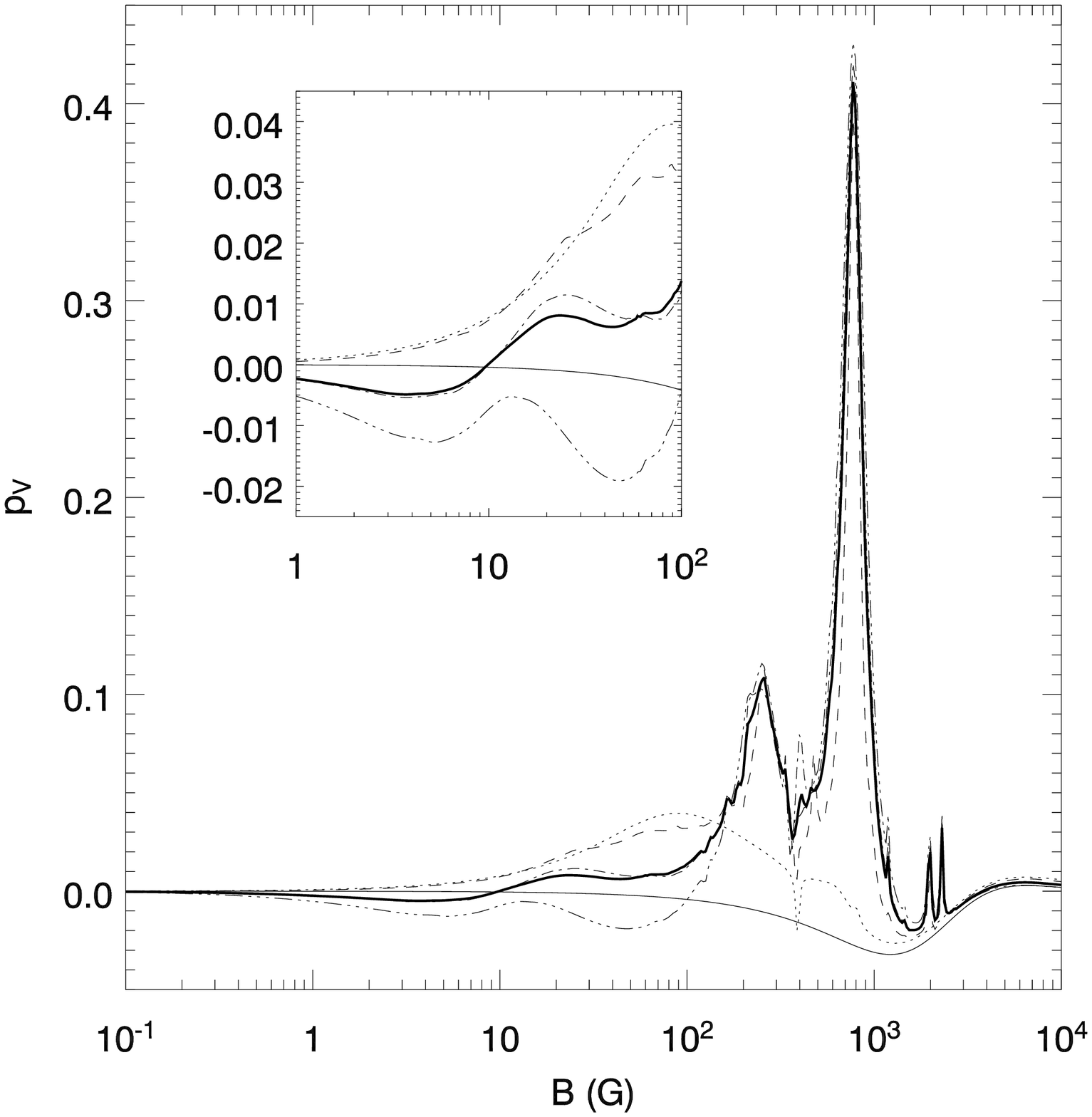}
\caption{\label{fig:Bvert.cir} 
Broadband, fractional circular polarization of the Ly$\alpha$ 
(left panel) and H$\alpha$ (right panel) radiation in a 
forward scattering event, as a function of the strength of a magnetic 
field directed along the line-of-sight, and in the additional presence 
of a randomly oriented electric field with $E=10\,\rm V\,cm^{-1}$. The 
thin solid line corresponds to the solution without electric fields. 
The thick solid line corresponds to the case of isotropic electric
fields, while the discontinuous lines show the contributions of
azimuth-averaged electric fields of various inclinations $\vartheta_\vE$.
Because of the symmetry properties of the scattering process, the 
curves for $\vartheta_\vE$ and $180^\circ-\vartheta_\vE$ are identical.
The insert in the right panel shows an enlargement of the plot of the 
net circular polarization of H$\alpha$ for vertical magnetic fields 
between $1\,\rm G$ and $10^2\,\rm G$.}
\end{figure}

\section{Scattering polarization in the presence of a deterministic 
magnetic field and a microturbulent electric field}
\label{sec:Eturbulent}

In this section we study the scattering polarization of hydrogen lines 
in the simultaneous presence of a deterministic magnetic field and a 
microturbulent electric field.

We first examine the case of forward scattering with the magnetic field 
directed along the line-of-sight (i.e., the $z$-axis of
Fig.~\ref{fig:geometry}). In this geometry, the symmetry considerations 
of the previous section still hold, because the magnetic field preserves
the cylindrical symmetry of the illumination process. So we can 
calculate the scattering polarization for a random-azimuth distribution 
of electric fields by averaging over the azimuth of the line-of-sight 
for a fixed azimuth of the electric field, using 
eq.~(\ref{eq:random_average}). 
%The final average over different 
%inclinations of the electric fields with respect to the direction of 
%incident radiation was calculated using an 8-point Gaussian quadrature
Once again, the case of completely isotropic fields is calculated using 
an 8-point Gaussian quadrature of the solutions for random-azimuth fields 
at appropriate inclinations \cite{AS64}.

The presence of a magnetic field introduces a chirality in the system, 
which is evidenced by the generation of a non-vanishing amount of 
broadband circular polarization (see Fig.~\ref{fig:Bvert.cir}, 
thin solid line). Ultimately this is determined by the presence of 
atomic orientation in the hydrogen levels, which is generated by the 
alignment-to-orientation (A-O) conversion mechanism \cite{LE69,KMN84}. 

The additional presence of an isotropic electric field, as small as 
$10\,\rm V\,cm^{-1}$, has a dramatic effect on the net circular 
polarization (NCP), as illustrated by Fig.~\ref{fig:Bvert.cir} 
(thick solid line). That figure also shows the contributions of 
electric fields with different inclinations 
with respect to the magnetic field. Electric fields parallel to the magnetic 
field alter quantitatively the amount of NCP. However, electric 
fields not aligned with the macroscopic magnetic field make, in general, 
a qualitatively more diverse contribution to the NCP. This is due to the
very complex pattern of level crossings and anti-crossings that is
produced for crossed fields. In particular, we notice the remarkable 
resonance of the NCP of H$\alpha$ at $B\approx 800\,\rm G$. 

The reason why these structures are typically very broad (even in 
a logarithmic scale) is that the A-O mechanism takes place for a 
range of Larmor frequencies comparable to the fine-structure 
separation within an atomic term. In fact, the origin of the A-O 
mechanism is the interference between different $J$ levels 
rather than level crossing itself. (However, when a level crossing 
occurs in the regime of the A-O mechanism, a sharp resonance of 
the NCP can be produced on top of the main structure.) The additional 
presence of an electric field, inclined with respect to the magnetic 
field, produces a sub-fine structuring of the atomic term (in the form 
of level anti-crossings), which modulate the NCP with additional
structures of diverse width and amplitude.

The important effects of microturbulent electric fields on the NCP 
of H$\alpha$ must be taken into account when interpreting broadband 
filter observations of the circular polarization in this line. In fact, 
all too often the assumption is made that the longitudinal Zeeman 
effect, with its characteristic antisymmetric signature, is the only 
mechanism responsible for the observed signal. Then, a non-zero NCP 
is explained in terms of velocity gradients along the line-of-sight, 
or even as instrumental cross-talk that needs to be removed, thus 
missing out on an important physical aspect of hydrogen plasma 
spectroscopy.

\begin{figure}[tp]
\flushright \leavevmode
\includegraphics[scale=.4]{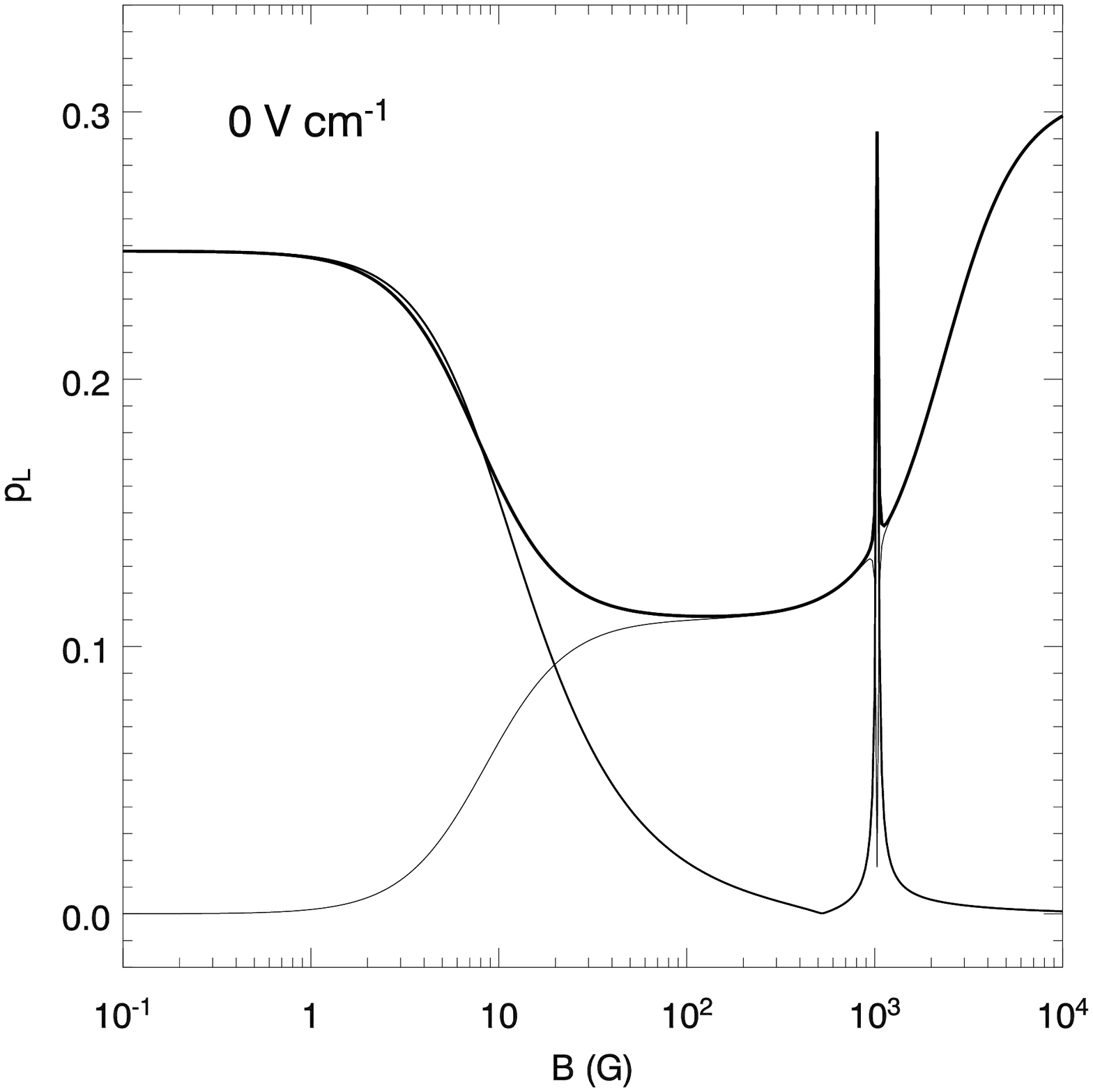}\hspace{5pt}
\includegraphics[scale=.4]{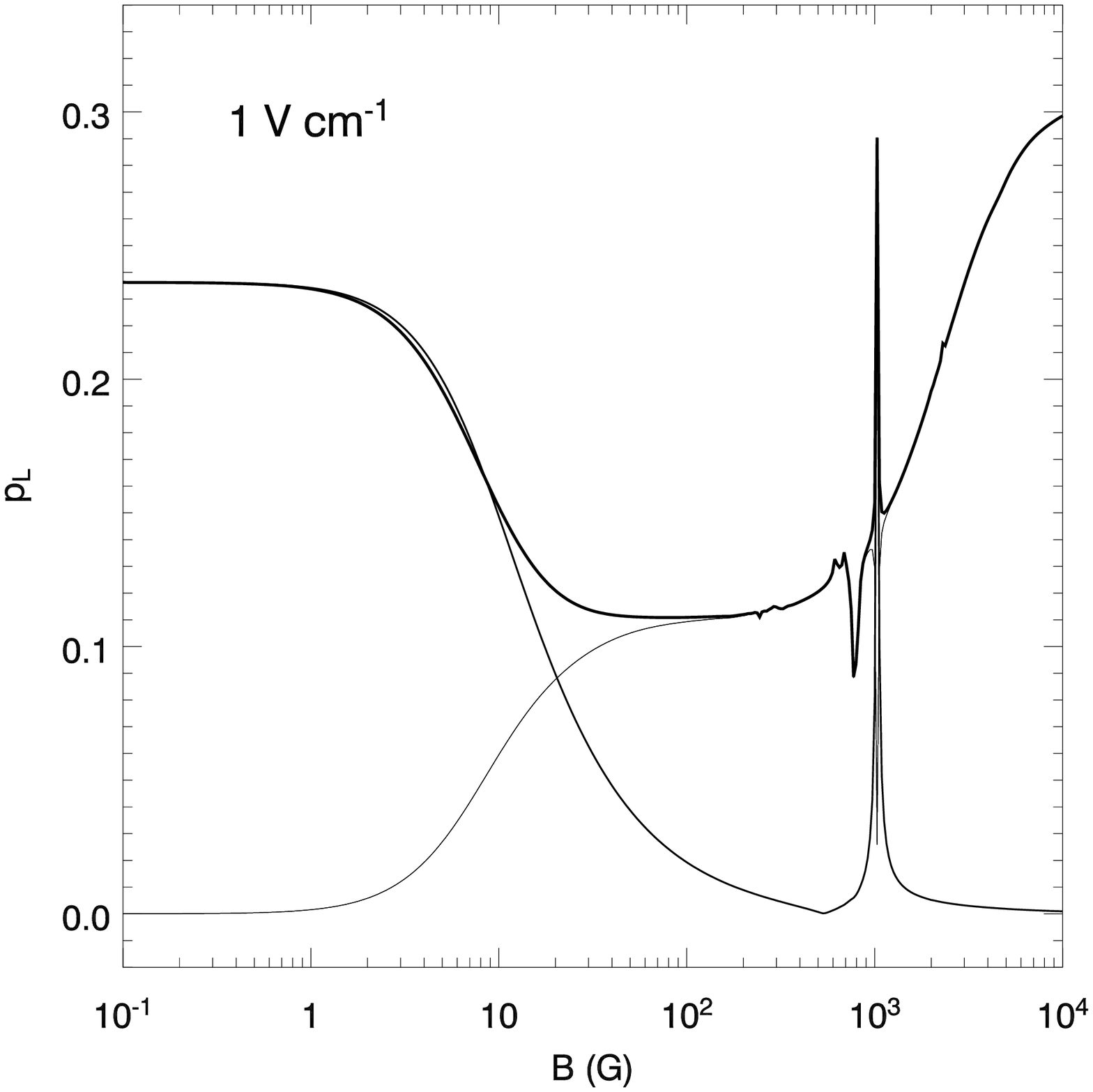}\vspace{10pt}
\includegraphics[scale=.4]{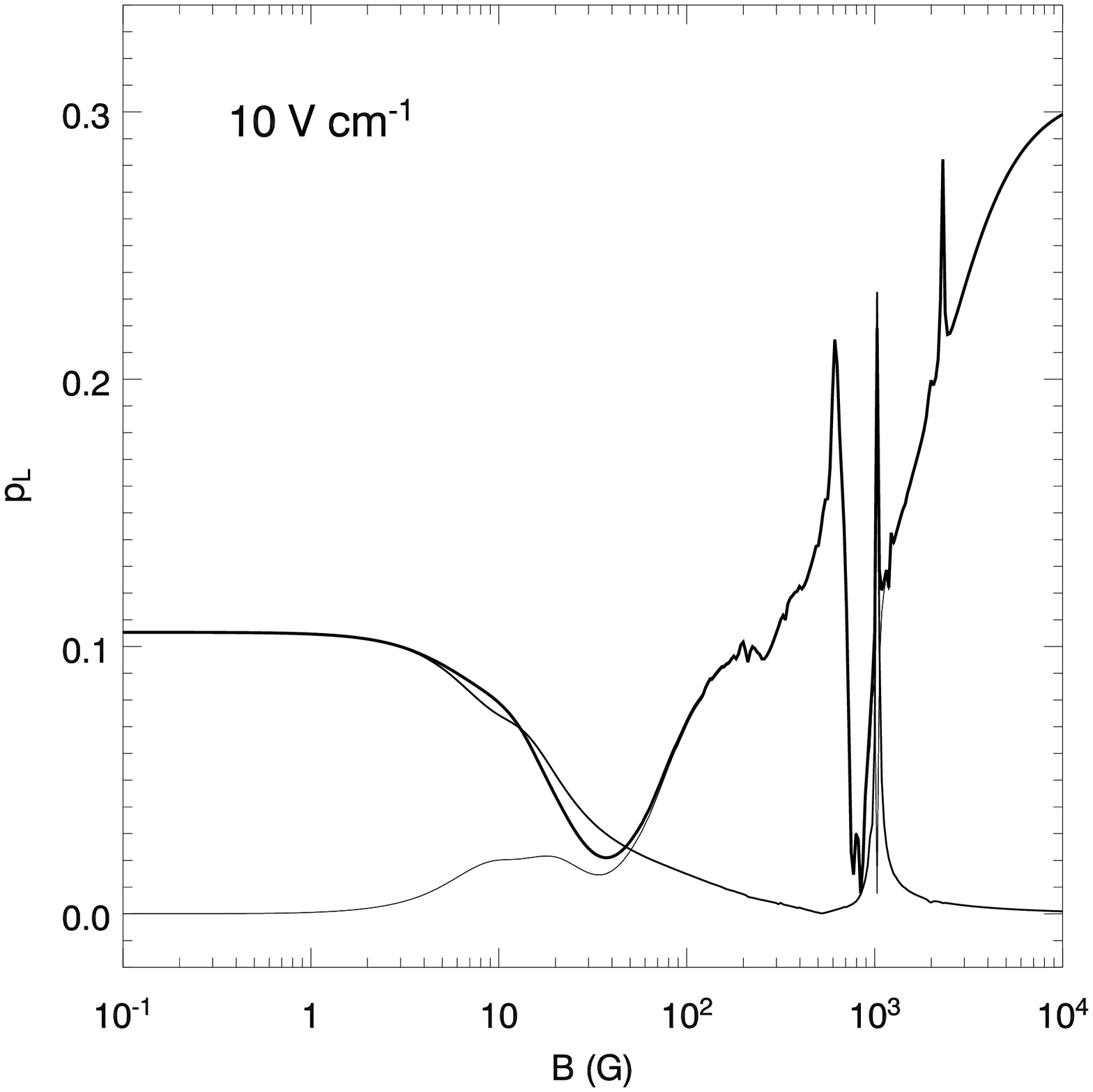}\hspace{5pt}
\includegraphics[scale=.4]{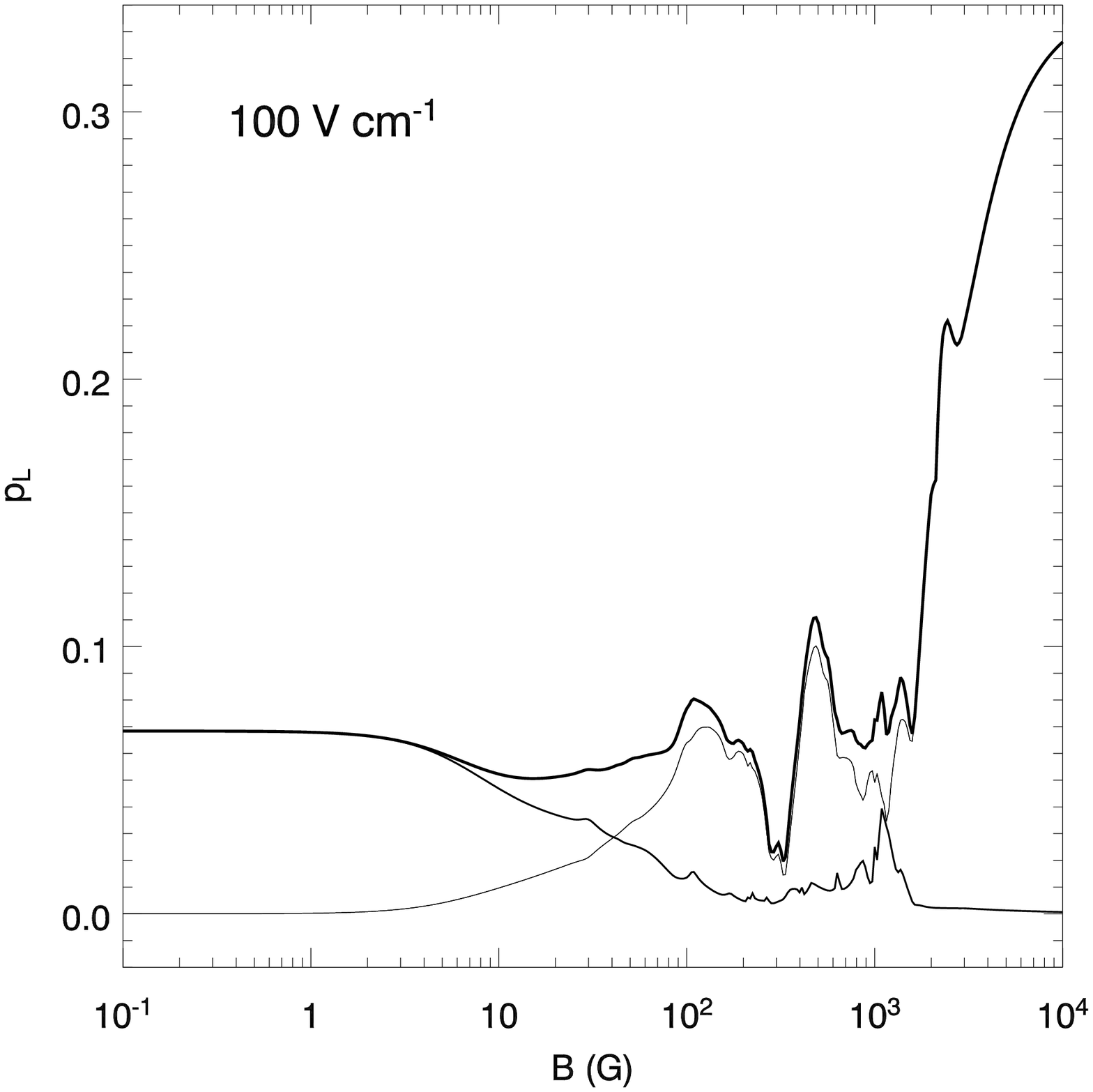}
\caption{\label{fig:Ha.Boriz.lin} %\small
Broadband, fractional linear polarization of the H$\alpha$ line 
of hydrogen in the presence of a horizontal magnetic field, and for
increasing strengths of a superimposed, isotropic electric field. 
The thin line corresponds to the case of forward scattering. The medium
line corresponds to the case of $90^\circ$ scattering, with the magnetic
field directed towards the observer. Finally, the thick line
corresponds to the case of $90^\circ$ scattering, with the magnetic
field at $90^\circ$ from the line-of-sight. We notice that the
``saturation'' level for strong magnetic fields, transverse to
the line-of-sight, is not constant for increasing electric strengths, as
a result of the competing regimes for the normal Zeeman effect and the
normal Stark effect.}
\end{figure}

\begin{figure}[t]
\flushright \leavevmode
\includegraphics[height=3in]{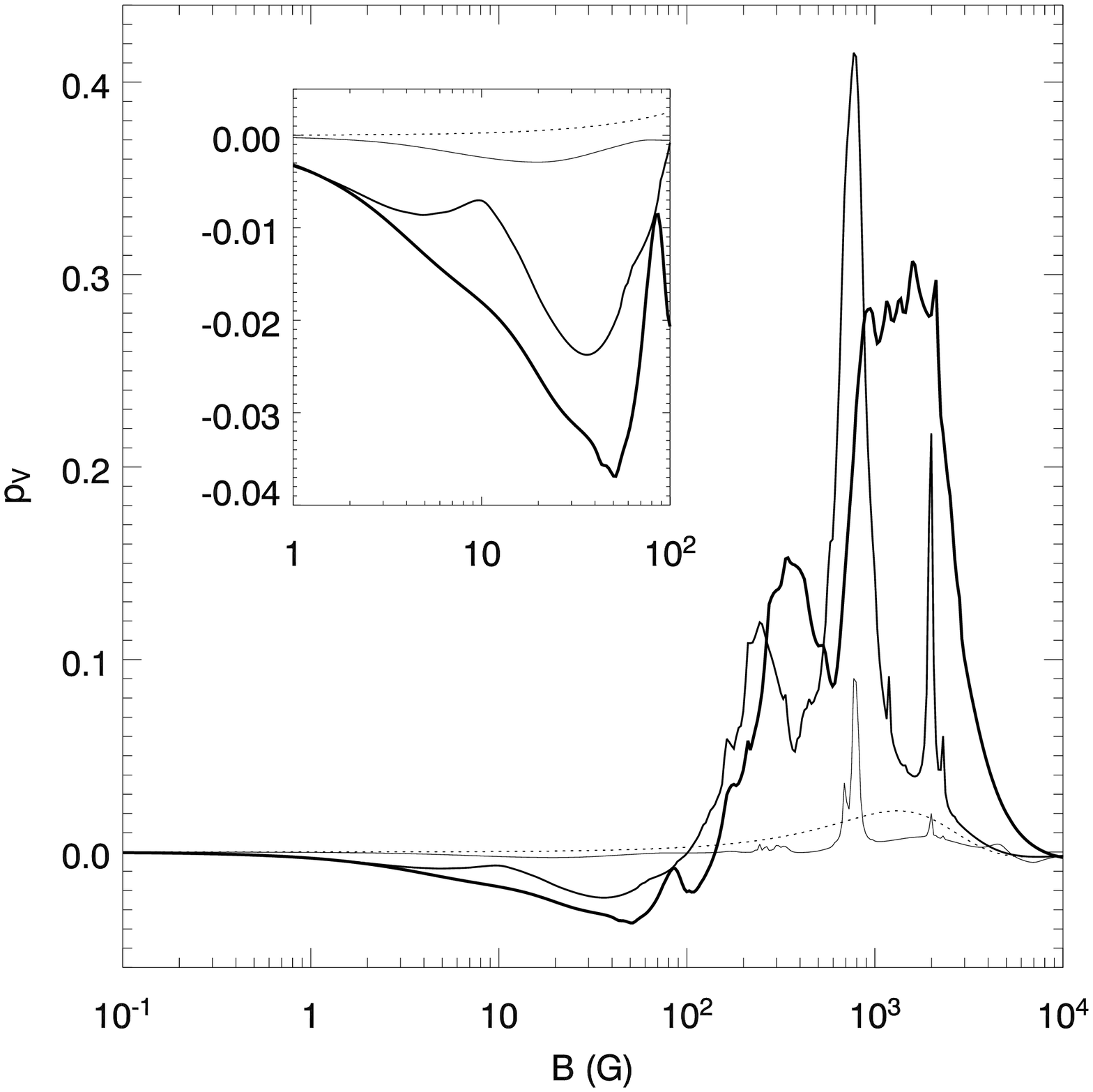}
\caption{\label{fig:Ha.Boriz.cir} %\small
Net circular polarization of the H$\alpha$ line of hydrogen in a 
$90^\circ$ scattering event, with a horizontal magnetic field directed 
towards the observer. Thin, medium, and thick solid lines correspond
to the cases in which an isotropic electric field of 1, 10, and
$10^2\,\rm V\,cm^{-1}$, respectively, is also present. The dotted line
represents the limit case with zero electric field. 
The insert shows an enlargement of the plot of the NCP of H$\alpha$ 
for horizonatal magnetic fields between $1\,\rm G$ and $10^2\,\rm G$.}
\end{figure}

Next we consider the case of a deterministic magnetic field oriented
at $90^\circ$ from the incident direction (horizontal field). 
We examine three fundamental scattering configurations for this 
case: a) forward scattering; b) $90^\circ$ scattering with the 
magnetic field directed along the line-of-sight; and c) $90^\circ$ 
scattering with the magnetic field oriented at $90^\circ$ from the
line-of-sight.
Because the magnetic field no longer preserves the cylindrical symmetry
of the scattering event, this time 
$\langle \varepsilon_i(\vk) \rangle_{(\vartheta_\vE,\varphi_\vE)}$ must be
calculated using a numerical angular quadrature over the whole sphere. For this
purpose, we adopted the 26-point quadrature given in \cite{AS64}.
%, \S 25.4.65. 
%It has eight points ($\pm 1/\sqrt{3}$, $\pm 1/\sqrt{3}$, $\pm 1/\sqrt{3}$)
%with weight 27/840; 12 points ($\pm 1/\sqrt{2}$, $\pm 1/\sqrt{2}$, 0),
%($\pm 1/\sqrt{2}$, 0, $\pm 1/\sqrt{2}$), (0, $\pm 1/\sqrt{2}$, 
%$\pm 1/\sqrt{2}$),
%with weight 32/840; and 6 points ($\pm 1$, 0, 0), (0, $\pm 1$, 0), 
%(0, 0, $\pm 1$) with weight 40/840. 

Figure \ref{fig:Ha.Boriz.lin} shows the broadband, fractional linear 
polarization, $p_L=\sqrt{p_Q^2+p_U^2}$, of the H$\alpha$ line of 
hydrogen for four different intensities (0, 1, 10, and $10^2\,\rm V\,cm^{-1}$) 
of the microturbulent electric field. The magnetic regime below
$10^2\,\rm G$ is dominated by scattering polarization, and its signal 
is an attenuated version of the limit case without electric fields
(compare the panels of Fig.~\ref{fig:Ha.Boriz.lin} for non-zero electric
fields, with the right panel of Fig.~\ref{fig:Balmer}).
%In the magnetic regime of strong fields the signal tends towards
%the same asymptotic limit. 
Instead, isotropic fields as weak as $10\,\rm V\,cm^{-1}$ significantly 
affect the linear polarization signal above $10^2\,\rm G$.

The strong resonance at $B\approx 10^3\,\rm G$ that is visible for $E\le
10\,\rm V\,cm^{-1}$ has already been discussed in the previous section,
and it occurs for $\omega_B=(4/9)\,\omega_{\rm FS}$, even in the absence
of electric fields. In the particular 
field geometry adopted for these figures, the other resonance 
at $\omega_B=(2/3)\,\omega_{\rm FS}$ is not visible. The addition of 
a microturbulent electric field as small as $10\,\rm V\,cm^{-1}$ 
modifies the energy structure of the hydrogen atom dramatically.
As a consequence, the pattern of resonances in the H$\alpha$ 
polarization is also affected. In particular, we notice the negative
resonance appearing around $B\approx 800\,\rm G$, which becomes more
important with the increasing electric strength. This resonance has 
the same origin of the dominant peak in the right panel of
Fig.~\ref{fig:Bvert.cir}, and it is due to an electric induced
anti-crossing between the $\rm 3\,{}^2P$ and $\rm 3\,{}^2D$ terms. 
The sub-fine structure separation determined by anti-crossing levels 
increases with the electric strength, eventually lifting 
the level crossing at $\omega_B=(4/9)\,\omega_{\rm FS}$. In the plots of
Fig.~\ref{fig:Ha.Boriz.lin}, this phenomenon is illustrated by the
increasing width of the resonance at $B\approx 800\,\rm G$, and the
leveling out of the peak at $B\approx 10^3\,\rm G$ for increasing
electric strengths. Other resonances are
also visible in those figures, appearing and disappearing as the energy
structure of the atom is modified by the increasing electric field.

It is also interesting to consider the NCP for this new magnetic 
configuration. Figure~\ref{fig:Ha.Boriz.cir} illustrates the case of 
the H$\alpha$ line. The dotted line corresponds to the purely magnetic 
limit, showing the NCP generated by the A-O mechanism, mostly 
between $10^2\,\rm G$ and $10^4\,\rm G$. 
It is notable the modification that microturbulent electric fields as 
weak as $1\,\rm V\,cm^{-1}$ already produce (thin solid line). In
particular, we notice once again the resonance at $B\approx 800\,\rm G$.
The effects determined by fields of 10 and $10^2\,\rm V\,cm^{-1}$ (medium 
and thick solid lines, respectively) are significantly larger, as
expected. 

%\begin{figure}[t]
%\plottwo{Ha.Bvert.cir.10.zoom.ps}{Ha.Boriz.cir.10.zoom.ps}{.1}
%\caption{\label{fig:Ha.cir.zoom} 
%Broadband, fractional circular polarization of the H$\alpha$ 
%line of hydrogen for a longitudinal magnetic field. The right panel 
%illustrates the case of forward scattering, and it is an enlargment 
%of the right panel of Fig.~\ref{fig:Bvert.cir} for magnetic fields 
%between 1 and $100\,\rm G$. The right panel illustrates the case
%of $90^\circ$ scattering, and it is an enlargement of
%Fig.~\ref{fig:Ha.Boriz.cir} for the same magnetic range. We notice the
%enhancement of the net circular polarization induced by a microturbulent
%electric field for magnetic fields that are typical of solar
%prominences.}
%\end{figure}

In concluding this discussion, it is important to remark that such 
microturbulent fields are characteristic of ionic densities between 
$10^{11}$ and $10^{13}\,\rm cm^{-3}$, which are common in many 
astrophysical plasmas, like stellar atmospheres (see Table~\ref{tab:table}). 
In particular, the inserts in the right panel of
Fig.~\ref{fig:Bvert.cir}, and in Fig.~\ref{fig:Ha.Boriz.cir}, show in 
details the modification of magnetic-induced circular polarization in 
the presence of microturbulent electric fields, for the restricted 
magnetic range 
$1\,{\rm G}\le B\le 10^2\,\rm G$. This is the regime of magnetic fields 
that can be expected in solar prominences, according to recent 
spectropolarimetric observations in the D$_3$ line of He~{\sc i} at
587.6 nm \cite{CA03} (which, to first order, is insensitive to 
electric fields). On the other hand, according to our current knowledge 
of the electron density in these structures 
($10^{10} \lesssim N_{\rm e} \lesssim 10^{12}$ cm$^{-3}$ \cite{TH95}), 
the average Holtsmark field that can be expected is of the order of 
$10\,\rm V\,cm^{-1}$ (see Table~\ref{tab:table}). We see that for 
typical fields in solar prominences (the average field of quiescent 
prominences is around $20\,\rm G$ \cite{CA03}), the presence of 
a microturbulent field characteristic of the electron density in those 
structures can induce a significant enhancement of the NCP of the 
H$\alpha$ line, which can easily exceed one order of magnitude, 
depending on the magnetic field strength and orientation.
This result could explain recent observations of anomalous NCP of 
H$\alpha$ in solar prominences
\cite{LA05}, which cannot be explained simply by invoking the
A-O mechanism in the presence of magnetic fields alone.

%In general, the average over the electric field realizations must be 
%done numerically. 
%Let $\rho^K_Q(JL, J'L')_{\{\vartheta_E\varphi_E\vartheta_B\varphi_B\}}$ be the 
%set of statistical tensor components for a configuration with an electric
%field along the $\vartheta_E$, $\varphi_E$ direction, 
%with an obvious notational shortening (shorhand?) of Eq.~(??):
%%and let's represent
%%the dependence of emissivity on the statistical tensors in Eq.~(??) as
%%$\varepsilon_i({\rm k})=\varepsilon[\rho^K_Q(JL,J'L')_{\{\vartheta_E\varphi_E\}}]$.
%%$\mathbb{\varepsilon}$
%\begin{equation}
%\frac{1}{4\pi}\oint\varepsilon_i(\boldsymbol{\hat{\rm k}}) 
%\sin\vartheta_E{\rm d}\vartheta_E {\rm d}\varphi_E \approx 
%\sum_{i j} w_{i j} \sum_{\alpha KQ} A_\alpha T^K_Q(i, \boldsymbol{\hat{\rm k}})
%\rho^K_Q(JL, J'L')_{\{\vartheta^i_E\varphi^j_E\vartheta_B\varphi_B\}},
%\end{equation}
%where $\{\vartheta^i \varphi^i\}$ and $\{w_{ij}\}$ are the nodes and weights
%of a discrete quadrature over the sphere, respectively.

It is important to understand in more detail the physical mechanism 
by which turbulent (as well as deterministic) electric fields are capable 
to induce a large degree of atomic orientation, when in the presence of 
a magnetic field, and why this can
happen for significantly smaller magnetic strengths than in the case of 
the magnetic-induced A-O mechanism. As noted above, such mechanism of 
\textit{electric re-orientation} is responsible for the appearance of 
a significant amount of NCP in the H$\alpha$ line, for magnetic strengths 
that are sufficiently small for the Zeeman-effect signal to be 
completely dominated by the intensity-like contribution of the atomic 
orientation. In \cite{CA05}, it was pointed out that this phenomenon can 
be traced back to the different pattern of level crossings and 
anti-crossings that is determined by the simultaneous presence of 
magnetic and electric fields. Here we want to make this argument more 
precise. The basic fact is that the A-O mechanism is efficient for 
Larmor frequencies of the order of the fine-structure separation 
between interfering levels. Because of the relatively large size of 
the fine structure in the lowest levels of hydrogen, this condition 
requires magnetic strengths in excess of $10^3\,\rm G$. When an 
electric field is also present (whether deterministic or turbulent), 
quantum interferences can arise between different $LS$ terms belonging 
to the same Bohr level. For example, the two levels ${}^2{\rm P}_{3/2}$ 
and ${}^2{\rm D}_{3/2}$ in the upper level of H$\alpha$ interfere with 
each other in the presence of an electric field, and their separation 
(Lamb shift) is about two orders of magnitude smaller than the 
fine-structure separation within each term. For this reason, the A-O 
mechanism becomes already efficient for magnetic fields of the order of 
$10\,\rm G$. For those magnetic strengths, the polarization amplitude 
of the Zeeman effect is very small, and so the circular-polarization 
profile is dominated by the contribution of the atomic orientation 
\cite{CA05}.

\section{Conclusions}
\label{sec:concl}

In this paper, we conducted a numerical study of the effect of turbulent 
electric fields on the scattering polarization of hydrogen lines. 
We first considered the case of a non-magnetic plasma, and successively the 
case in which a deterministic magnetic field is also present. 
We presented results for the Ly$\alpha$ and H$\alpha$ lines, which are
the most relevant for laboratory and astrophysical plasma studies. 
However, our general conclusions apply as well to all other transitions 
within the model atom used for this work (Ly$\alpha$, Ly$\beta$, Ly$\gamma$,
H$\alpha$, H$\beta$, Paschen~$\alpha$). 
%Although we 
%presented results only for the Ly$\alpha$ and H$\alpha$ lines---because of 
%their relevance for laboratory and astrophysical plasma studies---we 
%verified that the general conclusions of this work apply as well to all 
%other transitions within our model atom (which includes all Bohr levels 
%up to $n=4$). 

For the case of a non-magnetic plasma, we find that turbulent electric 
fields decrease the broadband linear polarization produced by scattering,
analogously to the depolarization of the scattered radiation produced by 
turbulent magnetic fields through the magnetic Hanle effect 
\cite{ST82,LL88,ST94,LL04}. This electric depolarization must always
occur in a hydrogen plasma, because of the ubiquitous, microscopically 
turbulent, electric fields produced by charged ions (mainly protons) 
surrounding the scatterer (Holtsmark field; see \cite{GR74} and
Table~\ref{tab:table}). However, it must be expected also in the presence 
of macroscopically turbulent electric fields, as far as the spatial 
scale of plasma turbulence is much smaller than the photon mean free 
path in the plasma. For the case of a Holtsmark field, the electric
depolarization provides an independent polarimetric tool for plasma 
density diagnostics. 

For the case of a magnetized plasma, we find that the magnetic scattering 
polarization is significantly modified by the additional presence of 
electric fields, even when these are isotropically distributed. In 
particular, we must expect a general decrease of the broadband, linear 
polarization by scattering in hydrogen lines that form in the presence of 
weak magnetic fields ($B\lesssim 10^2\,\rm G$; see
Fig.~\ref{fig:Ha.Boriz.lin}). Obviously, this additional 
depolarization induced by turbulent electric fields must be taken into 
account in the magnetic Hanle-effect diagnostics of hydrogen lines formed 
in the presence of either deterministic or turbulent magnetic fields.

We also find that turbulent electric fields can be responsible for a 
significant enhancement of the NCP of hydrogen lines, for particular 
magnetic field strengths. For example, the many-fold enhancement 
of the NCP of H$\alpha$ (see Fig.~\ref{fig:Bvert.cir} and
\ref{fig:Ha.Boriz.cir}), for gas densities typical of the solar 
chromosphere 
($N_{\rm e}\sim 10^{11}\,\rm cm^{-3}$; see Table~\ref{tab:table}), and 
for magnetic field strengths approximately between 1 and $10^2\,\rm G$, 
provides a very simple explanation of recent measurements of anomalous 
levels of NCP in H$\alpha$ observed in solar prominences \cite{LA05}.

The results presented in this paper are of diagnostic interest for 
polarization studies of magnetic plasmas in a large interval of 
electron densities. In particular, they also apply to 
quasi-equilibrium plasmas for which the magneto-hydrodynamic 
hypothesis excludes the possibility that deterministic electric fields 
can form in the reference frame of the magnetic sources. Therefore, 
the general conclusion of this investigation is that the effect of 
turbulent electric fields of statistical origin cannot be ignored in 
any polarimetric diagnostics of hydrogen lines in magnetized 
plasmas, because of the significant modification of the broadband
polarization that these electric fields can produce. In particular,
magnetogram techniques based on the observation of the H$\alpha$
line on the Sun should be carefully revised.

There are three principal limitations to our work. First of all, the
line formation theory adopted for this work does not include collisions.
It is possible that collisional depolarization might significantly
affect our results by decreasing the amount of broadband polarization
that can actually be achieved for a particular field geometry and
illumination condition. In particular, this might reconcile our
investigation with some polarimetric observations of H$\alpha$ that 
show only negligible amounts of broadband polarization.

A second limitation is that our results are obtained assuming 
an average strength for the turbulent electric fields (typically, 
the normal field strength of the Holtsmark theory), rather than a 
more physically appropriate distribution of field strengths. 
However, the results of Section~\ref{sec:turbulent} can be 
generalized directly by convolving the curves in the right panels of
Figs.~\ref{fig:Lyman} and \ref{fig:Balmer} with the appropriate 
distribution of electric strengths. The same 
can be done for the case of turbulent magnetic fields (left panels of 
Figs.~\ref{fig:Lyman} and \ref{fig:Balmer}), given the
distribution of magnetic strengths in the plasma. 
%More computationally challenging are the problems 
%studied in Section~\ref{sec:Eturbulent}, where one must necessarily 
%engage in demanding process of calculating the 
%weighted average (following the electric strength distribution) of many 
%realizations of the plots of 
%Figs.~\ref{fig:Bvert.cir}--\ref{fig:Ha.cir.zoom}.
%This choice was dictated exclusively by the limitation of computing 
%resources. The results of Section~\ref{sec:turbulent} can be
%generalized directly by convoluting the curves in the right panels of
%Figs.~\ref{fig:Lyman} and \ref{fig:Balmer} with the appropriate 
%distribution of electric strengths. Of course, the same generalization 
%can be done for the case of turbulent magnetic fields (left panels of 
%Figs.~\ref{fig:Lyman} and \ref{fig:Balmer}), if one has some knowledge 
%of the distribution of magnetic strengths in the plasma. 
For the problems studied in Section~\ref{sec:Eturbulent}, instead, one 
must necessarily engage in the more computationally intensive process 
of calculating the 
weighted average (following the electric strength distribution) of many 
realizations of the plots of 
Figs.~\ref{fig:Bvert.cir}--\ref{fig:Ha.Boriz.cir}.

The third simplifiying hypothesis of our work, which should be
investigated further, is the very assumption that the turbulent
electric fields be isotropic. This seems a reasonable hypothesis in 
the absence of other deterministic fields (whether magnetic or 
electric), like the cases considered in Section~\ref{sec:turbulent}. 
However, the additional presence of a deterministic field introduces 
a privileged direction, in response of which, for example, the ions in 
the plasma could organize themselves into anisotropic particle 
distributions, which ultimately would be responsible for anisotropic 
distributions of the perturbing electric fields \cite{SO73,DB76}. 
It would be very interesting to extend the methods illustrated in 
this paper to explore the possibility of detecting the 
spectro-polarimetric signature of such anisotropic distributions of 
ions in a plasma. 

%\textbf{
It must also be observed that the simple thermal agitation of 
the scattering atoms in a hot, magnetized plasma is directly 
responsible for the appearance of microturbulent electric fields of 
the $\bi{v}\times\bi{B}$ type, in the atomic rest frame. These, 
so-called \textit{motional}, electric fields add an anisotropic term 
(lying on the plane normal to $\bi{B}$) to the microturbulent fields 
due to the charged perturbers, even when the latter can be considered 
to be isotropically distributed in the plasma. In particular, this 
opens the intriguing possibility of a polarimetry-based diagnostics 
of plasma temperature.%}

%\textbf{
It is evident that a quantitative analysis of real observations of 
the scattering polarization in magnetized hydrogen plasmas must take 
into account all possible causes of anisotropy of the distribution of 
the microturbulent electric fields---like the ones discussed above---for 
a correct diagnostics of the magnetic field.%}

\end{document}